\documentclass[12pt]{article}
\usepackage{amsmath,amssymb,amsfonts,epsfig,graphicx}
\usepackage{amsmath,amssymb,epsfig}
\usepackage{cite}
\numberwithin{equation}{section}
\newcommand{\be}{\begin{equation}}
\newcommand{\bea}{\begin{eqnarray}}
\newcommand{\eea}{\end{eqnarray}}
\newcommand{\ba}{\begin{array}}
\newcommand{\ea}{\end{array}}
\newcommand{\ee}{\end{equation}}

\expandafter\ifx\csname mathbbm\endcsname\relax

\else

\newcommand{\bse}{\begin{subequations}}
\newcommand{\ese}{\end{subequations}}
\fi \textheight 22cm \textwidth 15cm \topmargin 1mm \oddsidemargin
5mm \evensidemargin 5mm

\begin{document}

\begin{titlepage}
\hfill
\vbox{
    \halign{#\hfil         \cr
           IPM/P-2005/081 \cr
                      } 
      }  
\vspace*{20mm}
\begin{center}
{\Large {\bf Semiclassical Strings Probing NS5 Brane Wrapped on $S^5$}\\
}

\vspace*{15mm}
\vspace*{1mm}
{Hajar Ebrahim\footnote{ebrahimi@theory.ipm.ac.ir}}
\vspace*{1cm}

{\it  Institute for Studies in Theoretical Physics
and Mathematics (IPM)\\
P.O. Box 19395-5531, Tehran, Iran \\ \vspace{3mm}}

\vspace*{1cm}
\end{center}

\begin{abstract}
We study little string theory on $R^1\times S^5$, defined by a
theory which lives on type IIA $N$ NS5 branes wrapped on $S^5$,
using its supergravity dual. In
particular we study semiclassical rotating closed strings in this
background. We also consider Penrose limit of this background that
leads to a plane wave on which string theory is exactly solvable.

\end{abstract}
\end{titlepage}
\section{Introduction}
Little string theory is a nonlocal theory that has many features
of string theory such as T-duality, though it is a
non-gravitational theory \cite{Berkooz:1997cq,s,hag}. It is the theory which
lives on the worldvolume of type IIA NS5 brane in its decoupling
limit. From holographic point of view little string theory on
$R^{1,5}$ is dual to type IIA string theory on linear dilaton
background \cite{Aharony:1998ub}. Linear dilaton background is the
one obtained by near horizon limit of NS5 brane. More precisely
the supergravity solution for $N$ NS5 branes is given by
\bea
ds^2&=&-dt^2+{d\vec{x}}^2+f(dr^2+r^2d\Omega_3^2),\cr
e^{2\phi}&=&{g_s}^2f\;\;\;,\;\;\;dB=2N{\alpha'}\epsilon_3\;\;\;,\;\;\;
f=1+\frac{N\alpha'}{r^2},
\eea
where $d\vec{x}$ parameterizes a 5-dimensional flat space and
$\epsilon_3$ is the volume of $d\Omega^3$. In the decoupling limit
where $g_s\rightarrow 0$ and $l_s$,$u=\frac{r}{g_s}$ are set to be
finite, the $N$ NS5 branes supergravity solution reads \bea
ds^2&=&-dt^2+{d\vec{x}}^2+\frac{N\alpha'}{u^2}(du^2+u^2d\Omega_3^2),\cr
e^{2\phi}&=&\frac{N\alpha'}{u^2}\;\;\;,\;\;\;dB=2N\alpha'\epsilon_3. \eea This
solution is called linear dilaton background which is conjectured
to be
 dual to little string theory on $R^{1,5}$. Therefore by making use of AdS/CFT
correspondence
\cite{{Maldacena:1997re},{Gubser:1998bc},{Witten:1998qj}} (for a
general review  see \cite{Aharony:1999ti}), one may use the
gravity dual side to understand some features of little string
theory side. Such studies have been done for the case where we
have little string theory on $R^{1,5}$ \cite{{Giveon:1999px},
{DeBoer:2003dd},
{Aharony:1999ks},{Minwalla:1999xi},{Aharony:1999dw}}. The
noncommutative deformation of this background and their
corresponding Penrose limits have also been studied in
\cite{{Alishahiha:1999ci}, {Harmark:2000ff},{Alishahiha:2000kw},
{Alishahiha:2000pu},{Mitra:2000wr},
{Oz:2002ku},{Alishahiha:2002zu},{Bhattacharya:2002zf},{Bhattacharya:2002qx},{Fuji:2002vs},{Kiritsis:2002kz},
{Hubeny:2002vf},{Alishahiha:2003xe}, {Matlock:2003dx}}.

Recently in the context of 1/2 BPS solutions \cite{Lin:2004nb} Lin
and Maldacena have obtained the supergravity solution of type IIA
$N$ NS5 branes wrapped on $S^5$, that is \cite{LM}
\begin{eqnarray}
 ds_{10}^{2} &=&N\alpha' \bigg{[} -2r \sqrt{\frac{I_{0}}{I_{2}}}dt^{2}+2r \sqrt{
\frac{I_{2}}{I_{0}}}d\Omega
_{5}^{2}+\sqrt{\frac{I_{2}}{I_{0}}}\frac{I_{0}}{ I_{1}}(dr
^{2}+d\theta ^{2})+\sqrt{\frac{I_{2}}{I_{0}}}\frac{
I_{0}I_{1}s^{2}}{I_{0}I_{2}s^{2}+I_{1}^{2}c^{2}}d{\Omega
}_{2}^{2}\bigg{]}  \cr B_{2} &=& N\alpha' \bigg{[}
\frac{-I_{1}^{2}cs}{I_{0}I_{2}s^{2}+I_{1}^{2}c^{2}}+\theta
\bigg{]}
 d^{2}\Omega  \cr
e^{\Phi } &=& g_0 N^{3/2} 2^{-1}\Bigg{(} \frac{I_{2}
}{I_{0}}\bigg{)} ^{\frac{3}{4}}\bigg{(}
\frac{I_{0}}{I_{1}}\bigg{)} ^{\frac{1}{2} }\bigg{(}
I_{0}I_{2}s^{2}+I_{1}^{2}c^{2}\bigg{)} ^{-\frac{1}{2}} \cr
C_{1}
&=&- {\alpha'}^{1/2}g_0^{-1} \frac{ 1}{ N } 4  \frac{I_{1}^2c}{I_{2}}
dt \cr
C_{3} &=&- {\alpha'}^{3/2}g_0^{-1}
\frac{4I_{0}I_{1}^{2}s^{3}}{I_{0}I_{2}s^{2}+I_{1}^{2}c^{2}}
dt\wedge d^{2}\Omega \label{first}\end{eqnarray} where ${I_n} (r)$
are a series of modified Bessel functions of the first kind. Also
$s$ and $c$ mean $\sin (\theta)$ and $\cos(\theta)$ respectively. This solution preserves 16 supercharges
 and has $R\times SO(3)\times SO(6)$ bosonic symmetry group. In our notation we have
 \bea
 d\Omega_5^2&=&d\theta_1^2+\cos^2\theta_1d\theta_2^2+\sin^2\theta_1d\Omega_3^2,\cr
d\Omega_2^2&=&d\phi_1^2+\cos^2\phi_1^2d\phi_2^2.
\eea
In spirit of AdS/CFT one may suspect that type IIA string theory on this new
 background is dual to little string theory on $R^1\times S^5$.
 If this is the case one should first check whether there is a notion of
 decoupling limit. This can be done by making use of scattering of graviton from
NS5 brane. Following \cite{Alishahiha:2000qf}
one can see that the scattering of transverse graviton will essentially lead to compute the scattering of a scalar field $\phi(r)$ the brane.
This scalar field will satisfy the Laplace
equation as follows \be
\partial_r(\sqrt{G}G^{rr}\partial_{r}\phi)+\partial_t(\sqrt{G}G^{tt}\partial_{t}\phi)=0,
\ee where $G$ is the determinant of the metric. Setting $\phi=
\alpha(r)\psi(r)e^{i\omega t}$ the above Laplace equation can be
recast to the following Schrodinger like  equation \be
{\partial_r}^2 \psi-V(r)=0, \ee where the potential is given by
\be
V(r)=\frac{1}{2}\frac{A''}{A}-\frac{1}{4}(\frac{A'}{A})^2-\frac{I_2}{2rI_1}\omega^2,
\ee with \be
A=r^3I_1^2\;,\;\;\;\;A'=\partial_rA\;,\;\;\;\;
A''=\partial^2_rA\;,\;\;\;\;\frac{\alpha'(r)}{\alpha(r)}=\frac{-A'}{2A}\;. \ee The
shape of the potential, having an infinite barrier, shows that the
supergravity solution (\ref{first}) is in the decoupling limit of NS5 branes,
the limit for which the modes in the throat will decouple from the
modes of the rest of the space. Therefore it is reasonable to say
that there is a little string theory which lives on $R^1\times
S^5$ that is dual to type IIA string theory on $N$ NS5 branes wrapped on $S^5$ background,
given by (\ref{first}). So we can use the gravity side to obtain
some information about little string theory side. For example one
could study the Penrose limit of this solution and also following
\cite{Gubser:2002tv} we could consider semiclassical rotating and
spinning closed string solutions on this background. This could
give us an insight of what the little string theory on $R^1\times
S^5$ might be.

The paper is organized as follows. In section 2, following the study
of localized rotating closed string in $S^2$, we will
study the Penrose limit of this background and we will see
that string theory on it, is exactly solvable. In section 3, we will study
semiclassical closed strings rotating and spinning in subsequently
$S^2$ and $S^5$. And the last section is devoted to conclusion.

\section{Plane Wave Limit}

As we said the background we are considering, (\ref{first}),
has $R\times SO(3)\times SO(6)$ bosonic symmetry. It means that
the isometries of the metric are time, related to a conserved
energy, three angle coordinates in $S^5$ which would lead to three spin
conserved charges, and one angle coordinate in $S^2$ which would
give a conserved angular momentum or R-symmetry charge. In the
following two sections, according to these isometries, we will
consider semiclassical closed strings that carry the corresponding
conserved charges to probe this background and understand it
better.

As the first case we will study a configuration in which the
semiclassical closed strings are centered around the origin of
$S^5$ sphere and stretched along radial coordinate, $r$. Also
these folded closed strings rotate in one direction in $S^2$. In
our notation the corresponding closed string configuration is
given by \bea
t=\kappa\tau\;,~~~r=r(\sigma)~,~~~\varphi_2=\nu\tau~,~~~\theta=\frac{\pi}{2}~,
\eea and all the other coordinates are set to zero. Using this
ansatz, the bosonic part of the superstring action is
\begin{equation}
S=\frac{-N}{4\pi}\int d\sigma d\tau
\bigg{(}2r\sqrt{\frac{I_0}{I_2}}\kappa^2+
\sqrt{\frac{I_2}{I_0}}\frac{I_0}{I_1}r'^2-\sqrt{\frac{I_2}{I_0}}\frac{I_1}{I_2}\nu^2\bigg{)},
\end{equation}
where $r'=\partial_\sigma r$. The solution or ansatz considered here should also satisfy the Virasoro condition that is
\begin{equation}
r'^2+(\frac{I_1^2}{I_0I_2}\nu^2-2r\frac{I_1}{I_2}\kappa^2)=0.
\end{equation}
The corresponding conserved charges are \be
E=\frac{N\kappa}{\pi}\int d\sigma
r\sqrt{\frac{I_0}{I_2}}\;,\;\;\;\;\;\;\;\;\;\;\;
J=\frac{N\nu}{2\pi}\int d\sigma
\sqrt{\frac{I_2}{I_0}}\frac{I_1}{I_2}. \ee Now the aim is to find
the dependence of energy, $E$, on angular momentum, $J$. This can
be done by making use of the Virasoro constraint. In general the
Virasoro constraint can be thought of "zero energy" condition for
a non-relativistic particle with a potential. We are looking for
periodic solutions that satisfy this equation. But noting the fact
that the first term, kinetic term, is positive definite, the
potential, $V(r)$, should be negative or zero and simultaneously
its derivative should be positive or zero.

In this ansatz the potential has no minimum or in other words the
potential is always negative and has negative slope. Therefore we
won't get any closed string solution except for $r=0$ that is
a zero size closed string solution. This will just happen
when we have $\nu=2\kappa$. Therefore in leading order, the relation between energy and angular momentum is given by
\begin{equation}
E=2J.
\end{equation}
What we have obtained are the states in which $E-2J$ is finite.
Now if we consider fluctuations around this classical solution, it
would lead to the Penrose limit of the background. The idea is to
consider the trajectory of a particle that is moving very fast
along one direction in $S^2$ and to focus on the geometry that
this particle sees. These fluctuations are
\begin{eqnarray}
t&=&\kappa\tau+\frac{\tilde{t}}{2^{1/4}\sqrt{N}}~,~~~\varphi_2=2\kappa\tau+\frac{\tilde{\varphi_2}}{2^{1/4}\sqrt{N}}~,~~~
r=\frac{2^{1/4}\tilde{r}}{\sqrt{N}}~,\cr
\theta&=&\frac{\pi}{2}+\frac{2^{1/4}y}{\sqrt{N}}~,~~~~~~~~\varphi_1=\frac{z}{2^{1/4}\sqrt{N}}
~,~~~~~~~~~~~\Omega=\tilde{\Omega}~.
\label{fluc}\end{eqnarray} We have also imposed the rescaling of
coordinates in these relations to get the leading finite terms in
the sigma model action when $N$ is large.

The bosonic part of the string sigma model action, in general
background, is \be S=\frac{1}{4\pi\alpha'}\int d^2\sigma
\sqrt{g}\bigg{[}\bigg{(}g^{ab}G_{\mu\nu}(X)+\epsilon^{ab}B_{\mu\nu}(X)
\bigg{)}\partial_aX^\mu\partial_bX^\nu+\frac{1}{2}\alpha'\Phi
R^{(2)}\bigg{]}, \ee
where $R^{(2)}$ is worldsheet curvature and
since in our case the metric is $\eta^{ab}$, $R^{(2)}$ is zero.
Using the fluctuations (\ref{fluc}) the first term of the
string action for the metric (\ref{first}) reads
 \bea
S_1&=&\frac{-1}{4\pi}\int d\sigma d\tau
\bigg{[}\partial_ay\partial^ay +\partial_az\partial^az+
\partial_aX_i\partial^aX^i
+\kappa^2(X_iX^i+16y^2+4z^2)\cr
&&\;\;\;\;\;\;\;\;\;\;\;\;\;\;\;\;\;\;\;\;\;\;
-4\partial_a\tilde{t}\partial^a\tilde{t}+\partial_a
\tilde{\varphi}_2\partial^a\tilde{\varphi}_2\bigg{]},
\eea
where
the angular coordinates in $S^5$ and radial coordinate, $r$, are
shown by $X_i$ where $i=1,...,6$ (6-dimensional flat space). On the other hand using the
definition of $B$ field in (\ref{first}) the second term of the
action would be \be S_2=\frac{-1}{2\pi}\int d\sigma d\tau
\;3\kappa y
\partial_\sigma z. \ee
Let us define the lightcone coordinates as
\begin{equation}
x^+=t+\frac{\varphi_2}{2}~,~~~~~~~x^-=t-\frac{\varphi_2}{2},
\end{equation}
therefore \be
x^+=2\kappa\tau+\frac{1}{2^{1/4}\sqrt{N}}(\tilde{t}+\frac{1}{2}\tilde{\varphi}_2)~,~~~~~x^-=
\frac{1}{2^{1/4}\sqrt{N}}(\tilde{t}-\frac{1}{2}\tilde{\varphi}_2).
\ee If we perform the following rescaling for $x^-$
\begin{equation}
x^-=\frac{\tilde{x}^-}{\sqrt{2}N}.
\end{equation}
and use the fact that $\kappa=\frac{1}{2}\partial_\tau x^+$,
lightcone gauge, the sigma model action is written as
\begin{eqnarray}
S&=&\frac{-1}{4\pi}\int d\sigma d\tau
\bigg{[}\partial_aX_i\partial^aX^i
+\frac{1}{4}\partial_\tau x^+\partial_\tau
x^+(X_iX^i+16y^2+4z^2)\cr
&&\;\;\;\;\;\;\;\;\;\;\;\;+\partial_ay\partial^ay+\partial_az\partial^az
-4\partial_ax^+\partial^a\tilde{x}^-+3y\partial_\tau
x^+\partial_\sigma z\bigg{]}.
\end{eqnarray}
This action is the one for eight massive bosons, related to eight
transverse coordinates. Therefore by using it we can see that the
fluctuations around closed zero size strings will see a plane wave
background given by the following metric \be
ds^2=\alpha'\bigg{[}dy^2+dz^2+dX_6^2
-\frac{1}{4}(X_6^2+16y^2+4z^2){dx^+}^2-4dx^+d\tilde{x}^-\bigg{]},
\label{pp}\ee and $B$ field
\begin{equation}
B_{+z}=3\alpha'y.
\end{equation}
Under this change of coordinates dilaton field will remain finite
if we rescale $g_0$ by $N^{-\frac{3}{2}}$. Also the RR field strengths corresponding to
$C_1$ and $C_3$ vanish when $N\rightarrow \infty$.

By making use of the action we can write the equations of motion for the strings
moving on this plane wave background as
\bea
\partial_a\partial^ay-3\kappa\partial_\sigma
z-16\kappa^2y&=&0\;,\;\;\;\;\partial_a\partial^az+3\kappa\partial_\sigma
y-4\kappa^2z=0\cr
\partial_a\partial^aX_i-\kappa^2X_i&=&0\;,\;\;\;\;i=1,..,6.\eea
These equations of motion are similar to what we have for harmonic
oscillator, six independent and two coupled ones. Therefore string
theory is exactly solvable on the plane wave background
(\ref{pp}). The results for normal oscillating modes are \be
\omega_{n}=\kappa\sqrt{1+\frac{n^2}{\kappa^2}}\;,\;\;\;\;\;\;\;\;\;\;\;\omega^\pm=\kappa\sqrt{10+\frac{n^2}{\kappa^2}\pm
3 \sqrt{4+\frac{n^2}{\kappa^2}}}. \label{osc}\ee Now we are able
to obtain energy, $E$, and angular momentum, $J$, relations for
the strings moving in this plane wave background. For the
classical closed string case we had $E=2J$. But when we consider
quantum fluctuations it would be \be
E-2J=\frac{1}{2\pi}\int d\sigma
\bigg{(}\kappa(X_iX^i+16y^2+4z^2)+4\sqrt{N}2^\frac{1}{4}\partial_\tau
\tilde{x}^-+3y\partial_\sigma z\bigg{)}. \label{e-2j}\ee
On the other hand we can write the Virasoro condition when we
consider these fluctuations, that is \bea
&\;&(\partial_\tau X_i\partial_\tau X^i+\partial_\sigma
X_i\partial_\sigma X^i)+(\partial_\tau y\partial_\tau
y+\partial_\sigma y\partial_\sigma y)+(\partial_\tau
z\partial_\tau z+\partial_\sigma z\partial_\sigma z)\cr
&\;&-\kappa^2(X_iX^i+16y^2+4z^2)-4(\partial_\tau
\tilde{t}\partial_\tau \tilde{t}+\partial_\sigma
\tilde{t}\partial_\sigma \tilde{t})+(\partial_\tau
\tilde{\varphi_2}\partial_\tau \tilde{\varphi_2}+\partial_\sigma
\tilde{\varphi_2}\partial_\sigma \tilde{\varphi_2})\cr
&\;&-8\kappa 2^{\frac{1}{4}}\sqrt{N}\partial_\tau\tilde{x}^-=0
\eea Using this Virasoro condition we can replace the term,
$8\kappa 2^{\frac{1}{4}}\sqrt{N}\partial_\tau\tilde{x}^-$, in
(\ref{e-2j}). The result is \be E-2J=H_{Light Cone}, \ee where
$H_{L.C.}$ is the Hamiltonian of the system in light cone gauge
and is given by \bea E-2J&=&\frac{1}{4\pi\kappa}\int
d\sigma \bigg{(}(\partial_\tau X_i\partial_\tau
X^i+\partial_\sigma X_i\partial_\sigma X^i)
+(\partial_\tau y\partial_\tau y+\partial_\sigma
y\partial_\sigma y)\cr
 &+&(\partial_\tau
z\partial_\tau z+\partial_\sigma z\partial_\sigma z)
+\kappa^2(X_iX^i+16y^2+4z^2)-4(\partial_\tau
\tilde{t}\partial_\tau \tilde{t}+\partial_\sigma
\tilde{t}\partial_\sigma \tilde{t})\cr &+&(\partial_\tau
\tilde{\varphi_2}\partial_\tau \tilde{\varphi_2}+\partial_\sigma
\tilde{\varphi_2}\partial_\sigma
\tilde{\varphi_2})+6\kappa y\partial_\sigma z\bigg{)}.
\eea Therefore by making use of (\ref{osc}) we have \bea
E-2J&\!=\!&\sum_n
N^{(6)}_n\sqrt{1+\frac{8N^2n^2}{J^2}}+N_n^{+}
\sqrt{10+\frac{8N^2n^2}{J^2}+ 3
\sqrt{4+\frac{8N^2n^2}{J^2}}}\cr &&\cr
&&\;\;\;\;\;\;+N_n^{-} \sqrt{10+\frac{8N^2n^2}{J^2}- 3
\sqrt{4+\frac{8N^2n^2}{J^2}}}, \label{e}\eea where
$N^{\pm}_n$ is the occupation number along $z$ and $y$ and also
$N^i_{n}$ are occupation numbers along six dimensional flat space,
characterized by $X_i$. In the classical case we saw that the
closed string satisfied the relation, $E-2J=0$. But now we can see
this relation will get some corrections which is controlled by
$\frac{N^2}{J^2}$.

Using AdS/CFT correspondence, there should be some operators in
little string theory on $R^1\times S^5$, for which both $E$ and
$J$ are large but $E-2J$ is finite. These states are not BPS,
because $E-2J$ gets quantum corrections, though these corrections
are under control and in fact the expansion parameter is given by
$\frac{N^2}{J^2}$. Note also that $\frac{N^2}{J^2}$ remains finite
in the $N\rightarrow\infty$ limit. To summarize one may conjecture
that type IIA string theory on plane wave background (\ref{pp}),
is dual to operators in little string theory on $R^1\times S^5$
whose $E$ and $J$ are large but $E-2J$ is finite and is given by
(\ref{e}).

\section{Rotating and Spinning Closed Strings}

In this section we will consider semiclassical closed strings
stretched along radius and rotate in $S^5$ and $S^2$, each in one
direction. The ansatz that describes this would be \bea
t=\kappa\tau~,~~~r=r(\sigma)~,~~~\varphi_2=\nu\tau~,~~~\theta_2=\omega\tau~,~~~\theta=\frac{\pi}{2}~,
\eea and other coordinates are set to zero. For this case the
action is written as
\begin{equation}
S=\frac{-N}{4\pi}\int d\sigma d\tau
(2r\sqrt{\frac{I_0}{I_2}}\kappa^2+\sqrt{\frac{I_2}{I_0}}\frac{I_0}{I_1}r'^2-2r\sqrt{\frac{I_2}{I_0}}\omega^2-\sqrt{\frac{I_2}{I_0}}\frac{I_1}{I_2}\nu^2).
\end{equation}
This ansatz will also satisfy the Virasoro condition that is
\begin{equation}
r'^2-\frac{r^2}{4}\frac{I_0-I_2}{I_0I_2}(4\kappa^2-\nu^2)\bigg{[}I_0-I_2\frac{4\omega^2-\nu^2}{4\kappa^2-\nu^2}\bigg{]}=0.
\label{diffr}\end{equation} The isometries of the action will
result in the following conserved charges \be
E=\frac{N\kappa}{\pi}\int d\sigma
r\sqrt{\frac{I_0}{I_2}}\;,\;\;\;\;\;
S=\frac{N\omega}{\pi}\int d\sigma
r\sqrt{\frac{I_2}{I_0}}\;,\;\;\;\;\;
J=\frac{N\nu}{2\pi}\int d\sigma
\sqrt{\frac{I_2}{I_0}}\frac{I_1}{I_2}.
 \label{esj}\ee
The aim of probing the background using strings is to obtain the
dependence of $E$ on $S$ and $J$, for generic $\kappa$, $\omega$
and $\nu$. In the first step, using the relations written in above
equation we will have \be
 E=\frac{4\kappa}{\nu}J+\frac{\kappa}{\omega}S.
 \label{E}\ee
To see whether this choice has a solution or not, we can use
 Virasoro
constraint.
 As what was done before, by looking at
the potential we see that the periodicity condition will be
satisfied if we consider $2\kappa
> \nu$ and $2\omega > \nu$. The turning point is $r_0$ and is
given by \be
\frac{I_0(r_0)}{I_2(r_0)}=\frac{4\omega^2-\nu^2}{4\kappa^2-\nu^2}.
\ee It is very difficult to obtain solutions precisely. So we will
study its long and short string limits. If we define \be
1+\eta=\frac{4\omega^2-\nu^2}{4\kappa^2-\nu^2}, \ee therefore
$\eta\rightarrow 0$ and $\eta\rightarrow \infty$ correspond to
long, $r\rightarrow\infty$, and short, $r\rightarrow 0$, limits,
respectively. Now we are ready to study short and long closed
strings rotating in this supergravity background.

\subsection{Short Strings}
In the short string limit, $r_0\rightarrow 0$, we can check that
whether there is a periodic solution. To see this we expand
(\ref{diffr}) for small $r\rightarrow 0$ we get \be
 {r'}^2\approx2(4\kappa^2-\nu^2)-\frac{1}{4}(4\kappa^2-\nu^2)(\frac{2}{3}+\eta)r^2,
 \ee
 and this equation will be satisfied by a periodic solution for
$r$ that is $r=r_0\sin\sigma$, if we have
 \be r_0=\sqrt{2(4\kappa^2-\nu^2)}\;,\;
 \;\;\frac{1}{4}(4\kappa^2-\nu^2)(\frac{2}{3}+\eta)=1.
 \ee
 Therefore using the definition of $\eta$ and the fact that in short string case
 $\eta\rightarrow\infty$ we obtain
\be 4\kappa^2-\nu^2\sim \frac{4}{\eta}\;,\;\;\;\;\;
4\omega^2-\nu^2\sim \frac{4}{\eta}+\eta. \label{kw}\ee Now using
the relation obtained for angular momentum $J$ and expanding it
for $r\rightarrow 0$, the dependence of $J$ on $\nu$ will be \be
J\sim\sqrt{2}N\nu. \ee Also if we do the same for
$S$ we get \be S\sim\frac{2\sqrt{2}N\omega}{\eta}.\ee
Plugging this equation into (\ref{kw}) we get an expression for
$\omega$ at leading order, which can be used to obtain
$\frac{1}{\eta}$ as follows \be
\frac{1}{\eta}\sim\frac{S}{\sqrt{2}N\sqrt{4+\frac{J^2}{2N^2}}}.\ee
We can substitute the above equation in (\ref{kw}) and find the
final results for $\kappa$ and $\omega$ as
  \bea
4\kappa^2&\sim&\frac{J^2}{2N^2}+\frac{4S}{\sqrt{2}N}\frac{1}{\sqrt{4+\frac{J^2}{2N^2}}},\\
4\omega^2&\sim&4+\frac{J^2}{2N^2}+\frac{2\sqrt{2}S}{N}\frac{1}{\sqrt{4+\frac{J^2}{2N^2}}}.
\eea We need to find the dependence of $E$ on $S$ and $J$. Using
(\ref{E}) and the relations for $\omega$, $\nu$ and $\kappa$ the
final result will be \be
E\approx\sqrt{\frac{2J^2}{N^2}+\frac{16\sqrt{2}S}{N\sqrt{16+\frac{2J^2}{N^2}}}}
\bigg{(}\sqrt{2}N+\frac{S}{\sqrt{16+\frac{2J^2}{N^2}
+\frac{16\sqrt{2}S}{N\sqrt{16+\frac{2J^2}{N^2}}}}}\bigg{)}.\ee For
the case where both $S$ and $J$ are small we get \be
E^2\approx8\sqrt{2}NS+4(J^2+S^2),\ee which actually represent
Regge trajectories in the flat space. The other limit is $S\gg J$
for which energy is given by \be E\approx
S-\frac{\sqrt{2}}{16N}J^2-\sqrt{2}N+\sqrt{S}\bigg{(}2^{\frac{7}{4}}\sqrt{N}-
\frac{1}{16}({\frac{\sqrt{2}}{N}})^{\frac{3}{2}}J^2\bigg{)}.\ee
Also when $S\ll J$ we can find energy as \be E\approx
2J+S+\frac{4N^2S}{J^2}-\frac{8N^4S}{J^4}.\ee It is easily seen
that this expression for energy is an expansion in terms of
$\frac{N^2}{J^2}$ and is related to the leading quantum term in
the spectrum of string on plane wave background, (\ref{e}).
\subsection{Long Strings}
As we saw the long string limit corresponds to
$r_0\rightarrow\infty$ and $\eta\rightarrow 0$. In (\ref{diffr})
we expand ${r'}^2$ for $r\rightarrow\infty$, therefore we find \be
{r'}^2\approx(3\kappa^2-\nu^2+\omega^2)+2r(\kappa^2-\omega^2). \ee
Considering the fact that $r_0$ is the turning point $(r'=0)$, we
obtain \be r_0=\frac{2}{\eta}+\frac{1}{2}. \ee On the other hand
we note that $r'=\frac{dr}{d\sigma}$ and also for
$0<\sigma<\frac{\pi}{4}$ the function $r(\sigma)$ increases from
zero to a maximal value $r_0$, so \be
2\pi=\int_0^{2\pi}d\sigma\approx\frac{4}{\sqrt{2(\omega^2-\kappa^2)}}\int_0^{r_0}\frac{dr}{\sqrt{\frac{2}{\eta}+\frac{1}{2}-r}}.
\ee Therefore we get \be
\omega^2-\kappa^2\sim\frac{4}{\pi^2}(\frac{4}{\eta}+1),
\label{wk}\ee and using the definition of $\eta$ we find \be
4\kappa^2-\nu^2\sim\frac{16}{\pi^2\eta}(\frac{4}{\eta}+1).\label{kv}\ee
Now if we expand the definition of $J$ and $S$ for large $r$ we
get that \be
J\sim N\nu\;,\;\;\;\;\;S\sim\frac{2N\omega}{3}
(\frac{4}{\eta}-2),\label{js}\ee where we have used (\ref{esj}).
In this step we should try to obtain a relation for $\eta$
dependence on $S$ and $J$. To do this we use (\ref{wk}) and
(\ref{kv}) to find a relation for $\omega$ in terms of $\eta$ and
$\nu$. Now if we substitute the result in the equation for $S$
(\ref{js}), we get a second order equation for $\eta$ that is \be
S+\frac{49N}{12\pi}-\frac{\pi J^2}{12N}-\frac{4N}{3\pi\eta}(\frac{8}{\eta}+1)\approx
0,\ee which can be solved to find
\be\frac{1}{\eta}\approx\frac{1}{16}(-1+\sqrt{99+\frac{24\pi S}{N}-\frac{2\pi^2J^2}{N^2}}).\ee
We substitute it in equations (\ref{wk}) and (\ref{kv}) and get
the final results for $\kappa$ and $\omega$ that are \bea
4\omega^2&\sim&\frac{J^2}{2N^2}+\frac{6S}{\pi
N}+\frac{36}{\pi^2}
+\frac{9}{2\pi^2}\bigg{(}\sqrt{99+\frac{24\pi S}{N}-\frac{2\pi^2J^2}{N^2}}\bigg{)},\\
4\kappa^2&\sim&\frac{J^2}{2N^2}+\frac{6S}{\pi
N}+\frac{24}{\pi^2}
+\frac{1}{2\pi^2}\bigg{(}\sqrt{99+\frac{24\pi S}{N}-\frac{2\pi^2J^2}{N^2}}\bigg{)}.\eea
Now we can use the equation obtained for energy, (\ref{E}), and
obtain energy in terms of angular momentum and spin. Energy in the
limit where $S$ is large and $J$ is small is given by\be E\approx
\frac{14N}{3\pi}+S+\frac{4}{3}\sqrt{\frac{6NS}{\pi}}+\frac{55\sqrt{6}}{36}{(\frac{N}
{\pi})}^{\frac{3}{2}}\frac{1}{\sqrt{S}}+\sqrt{\frac{\pi}{6N}}\frac{J^2}{\sqrt{S}}.\ee
We can see that in contrast to similar cases in $Ads_5\times S^5$
\cite{Gubser:2002tv,Frolov:2002av}, the corrections to energy are
not in terms of $\ln S$, but in terms of $\sqrt{S}$. Of course
this might be understood from the fact that in the $AdS_5\times
S^5$ case we are dealing with a gauge theory. It should also be
mentioned that limits in which $J$ is large or $J$ is greater than
$S$ in this long string limit are not reasonable.

\section{Conclusion}
In this paper we have studied the 1/2 BPS geometry of $N$ NS5
branes wrapped on $S^5$ using folded closed string probes. We have
first considered a point like closed string configuration which
rotates in $S^2$. The string sigma model expansion around this
classical solution resulted in pp wave limit of this background.
Such a result should be compared to pp wave limit of NS5 branes in
$R^{1,5}$ case. The pp wave metric we have obtained for $N$ NS5
branes wrapped on $S^5$, is 10-dimensional but for the case in
which NS5 brane is defined on $R^{1,5}$ the result is
4-dimensional plane wave times 6-dimensional flat space
\cite{Kiritsis:2002kz,Hubeny:2002vf,Alishahiha:2003xe,Matlock:2003dx}.

In the case we have considered, little string theory on $R\times
S^5$ is dual to type IIA string theory on $N$ NS5 branes wrapped
on $S^5$ background. Actually one may interpret the strings in
little string theory as D2 branes stretched between two NS5 branes
in type IIA string theory. In the limit where NS5 branes approach
each other we will have to deal with tensionless strings in little
string theory side. Now using the duality one may deduce that
little string theory on $R\times S^5$ is composed of closed
strings. Because on $S^5$ all dimensions are compact and therefore
periodic. This is in contrast with having just open strings in
little string theory on $R^{1,5}$.

We also considered short and long string limits of folded closed
strings, centered around the origin and stretched along the radial
coordinate. Such a configuration rotates along one direction in
$S^2$ and also spins along one direction in $S^5$. Such a
configuration would result in a state in little string theory that
has spin. It should be mentioned that such states do not exist in
little string theory on $R^{1,5}$. Because there are not such
classical closed string configurations in its supergravity dual.
In the short string case as expected we obtained Regge trajectory
as in the flat space when $S$ and $J$ are small. In long string
limit we have also seen that corrections to energy are in terms of
$\sqrt{S}$.

One could also consider the most general case in which the folded
closed string rotates in one direction in $S^2$ and spins in three
directions in $S^5$. Such a configuration that is called
multi-spin string solution, carries all the conserved charges
related to the isometries of the metric \cite{Frolov:2003qc}.

One may study the background by probing it with
NS5 brane wrapped on $S^5$ or D2 brane wrapped on $S^2$. Such
cases could be compared to giant gravitons in $AdS_5\times S^5$,
where two different D3 branes could wrap over two 3-spheres of
$AdS_5\times S^5$ in global coordinates. It would also be 
interesting to study the genuine dynamics of stings such as splitting
in this background. Such analysis has been done for $AdS_5\times S^5$ background, for example,
in \cite{{Peeters:2004pt},{Peeters:2005pb}}.

Finally we note that since the string theory on plane wave
background (\ref{pp}) is exactly solvable, one could also study
open string which would result in different D-brane configurations
that might exist in this background. In this paper we have only
studied the bosonic part of the string sigma model action. One may
also consider the fermionic part. Such study has been done in the
case of little string theory living on $R^{1,5}$
\cite{Matlock:2003dx}.

~~~~~~~~~~~~~

\noindent\textbf{Acknowledgments}

I would like to thank M. Alishahiha for bringing my attention
to this problem and for useful discussions. I would also like to
thank B. Chandrasekhar, A.E. Mosaffa and especially M.M. Sheikh-Jabbari for useful
discussions and comments.

\end{document}